\documentclass [aps,prb,twocolumn,showpacs,floatfix]{revtex4} 
\usepackage{graphicx}
\usepackage{dcolumn}% Align table columns on decimal point
\usepackage{bm}% bold math

\begin{document} 

\title{Static dielectric response and Born effective charge of BN nanotubes 
from {\it ab initio} finite electric field calculations}  
\author{G.Y. Guo$^{1,2,*}$, S. Ishibashi$^{1}$, T. Tamura$^{1}$, and K. Terakura$^{3}$}
%\author{G.Y. Guo$^{1,2}$\footnote{Electronic address:
%gyguo@phys.ntu.edu.tw}, S. Ishibashi$^{1}$, ???, and K. Terakura$^{3}$}
\address{$^{1}$Research Institute for Computational Sciences (RICS),
National Institute of Industrial Science and Technology (AIST), Tsukuba, 
Ibaraki 305-8568, Japan\\
$^{2}$Department of Physics and Center for Theoretical Sciences, National Taiwan University, Taipei 10617, Taiwan\\
$^{3}$Creative Research Institute "Sousei", Hokkaido University, Sapporo 001-0021, Japan}

\email{gyguo@phys.ntu.edu.tw}

\date{\today}

\begin{abstract}
{\it Ab initio} investigations of the full static dielectric response and
Born effective charge of BN nanotubes (BN-NTs) have been performed for the
first time using finite electric field method. It is found
that the ionic contribution to the static dielectric response of BN-NTs is
substantial and also that a pronounced chirality-dependent oscillation is superimposed
on the otherwise linear relation between the longitudinal electric polarizability
and the tube diameter ($D$), as for a thin dielectric cylinderical shell. 
In contrast, the transverse dielectric response of 
the BN-NTs resemble the behavior of a thin (non-ideal) conducting cylindrical shell
of a diameter of $D+4$\AA$ $, with a screening factor of 2 for the inner electric field. 
The medium principal component $Z_y^*$ of the Born effective charge corresponding to 
the transverse atomic displacement tangential to the BN-NT surface, has a pronounced 
$D$-dependence (but independent of chirality), while the large longitudinal 
component $Z_z^*$ exhibits a clear chirality dependence (but nearly $D$-independent), 
suggesting a powerful way to characterize the diameter and chirality of a BN-NT.

%It is also found that for each BN-NT, all the atoms of the same species have 
%the same three principal components of the Born effective charge with 
%the small and medium components having a pronounced monotonic $D$-dependence 
%(but independent of chirality) and the large componet 
%exhibiting a chirality dependence (but nearly $D$-independent), 
%suggesting a useful way to characterize a BN-NT.
%We report {\it ab initio} calculations of the static dielectric constant, electric 
%polarizability and
%Born effective charges of BN nanotubes (BN-NTs) as well as the single graphitic
%BN sheet and bulk BN structures, using local density approximation + finite-field
%electric-enthalpy theory.
%We find that the ionic contribution to the static dielectric constant and
%electric polarizability is substantial, though the full dielectric constant still
%has a weak dependence on diameter and chirality and the polarizability per unit length
%is roughly proportional to the tube diameter, as in previous independent-particle calculations. 
%We also find that for each BN-NT, all the atoms of the same species have three
%same principal Born effective charges with the large and small Born effective
%charges being nearly independent of diameter and chirality. The intermediate
%Born effective charge, however, has a pronounced linear dependence on diameter,
%indicating that the diameter of a BN-NT could be determined by measuring its
%Born effective charges.
\end{abstract}


\maketitle

\section{Introduction}

Since the discovery of carbon nanotubes (CNTs) 
in 1991~\cite{iij91}, carbon and other nanotubes have attracted 
considerable interest worldwide because of their unusual properties and great 
potentials for technological applications. For example, semiconducting
CNTs could function as nanoscale field-effect transistors \cite{tan98}.
It was also predicted that nanotori formed from metallic CNTs may 
exhibit giant paramagnetic moments.~\cite{liu02a} 
%Furthermore, chiral CNTs 
%are expected to exhibit a number of unusual optical properties such as 
%optical activity, circular dichroism and second harmonic 
%generation (see ~\cite{guo04} and references therein).
%Soon after the discovery of CNTs it became obvious that similar nanostructures
%could be formed by other elements and compounds which form layered structures
%bearing some resemblance to graphite. For example, hexagonal BN ($h$-BN) was
%predicted on the basis of theoretical calculations~\cite{rub94,bla94}
%to be capable of forming nanotubes, a prediction which was later confirmed
%experimentally by the synthesis of such nanotubes.~\cite{cho95} 
%Both single-walled and multiwalled BN nanotubes (BN-NT) can now be readily 
%synthesized.~\cite{lee01} 
Though CNTs continue to attract great interest, other nanotubes such as 
BN nanotubes (BN-NTs)\cite{cho95} are interesting in their own right and may 
offer different possibilities for technological applications
that CNTs cannot provide. In particular, as far as the optical and 
opto-electronic applications of nanotubes are concerned, BN-NTs could be
superior to CNTs because BN-NTs are uniformly insulating, independent of
their chirality. \cite{bla94} 
%Furthermore, BN-NTs tend to have a zigzag structure.~\cite{lee01}
%Though it is interesting that, depending on their chirality, 
%CNTs can be metallic or semiconducting or insulating~\cite{guo04}, it is still impossible
%to grow CNTs with a pre-specified chirality at present.
Furthermore, recent experiments indicate that BN-NTs exhibit stronger resistance
to oxidation at high temperatures than CNTs.~\cite{che04}
 
Therefore, the properties of both
single-walled and multiwalled BN-NTs have
been intensively studied experimentally and theoretically in recent years.
%(see, e.g., Ref. \onlinecite{guo05} and references therein). 
In this paper, we focus on the dielectric response, including
static dielectric constant, electric polarizability and Born effective
charge, of BN-NTs.
A detailed knowledge of dielectric response is needed, e.g., to
characterize optical excitations, screening at contacts, and plasmons in nanotube
arrays. Variations in size and chirality may or may not affect dielectric
properties, which in turn may be exploited to align nanotubes during, e.g.,
plasma-enhanced chemical vaposition synthesis \cite{zha01,yu01} and to separate
different tubes in solutions \cite{kru03}. Synthesis and separation of specific
nanotubes remain an important challenge.   

Previously, the static dielectric constant and electric polarizability
of nanotubes were typically obtained as the zero frequency limit of
the optical conductivity (see, e.g., Refs. \onlinecite{guo04,guo05} and
references therein), which is calculated using either semiempiral 
tight-binding method or {\it ab initio} local density approximation (LDA) 
or generalized gradient approximation (GGA) within independent particle 
linear response theory (IPLRT).
In this way, only the electronic contribution to the static dielectric
constant ($\epsilon_{\infty}$) is calculated.\cite{ben95,guo04,guo05,wir05} 
Furthermore, local
field effects, which can be substantial for the electric field applied
perpendicular to the tube axis,\cite{ben95}, are neglected.\cite{guo04,guo05}
Here, we apply the latest finite-field electric-enthalpy theory \cite{sou02,uma02}
to study the static dielectric properties of BN-NTs. This theory
allows us not only to obtain the full dielectric constants including
both electronic and ionic contributions but also to include the
local field effects. Furthermore, also using this theory, we perform
the first systematic study of the Born effective charge of the BN-NTs.
A knowledge of the dependence of Born effective charge on the
diameter and chirality of the nanotubes not only is important to understand
their dynamical properties but also would allow us to characterize
the nanotubes using, e.g., Raman and infrared spectroscopies. 
%Identication of if specific nanotubes is another great challenge. 

\section{Theory and computational details}

We use the latest finite field 
electric-enthalpy theory \cite{sou02,uma02} to
calculate field-induced electric polarization {\bf P} and atomic forces {\bf F}.
For a small finite field {\bf E}, the dielectric constant is
$\epsilon_{\alpha\beta} = \delta_{\alpha\beta} + 4\pi\chi_{\alpha\beta}$ where
$\chi_{\alpha\beta} = P_{\alpha}/E_{\beta}$. If the atoms are kept fixed, this yields
the electronic contribution $\epsilon_{\infty}$ (so-called clamped ion dielectric 
constant) and if both the electrons and atoms are 
allowed to relax in response to the field, the full static dielectric constant
$\epsilon_{static}$ is obtained. The electric polarizability $\alpha$ is given
by $\epsilon_{\alpha\beta} = \delta_{\alpha\beta} + 4\pi\alpha_{\alpha\beta}/\Omega$
where $\Omega$ is the unit cell volume.
%Note that so-called local field effects are included in this finite field theory.
%In contrast, local field effects are neglected in
%many previous {\it ab initio} calculations (see, e.g.,
%Refs. \onlinecite{guo04,guo05}), which are based on single-particle approximation.
%Local field effects are found to be rather significant for the transverse
%dielectric properties of nanotubes, though they are negligible for the longitudinal
%properties, as will be demonstrated below.
%Furthermore, only the electronic contribution $\epsilon_{\infty}$ is calculated
%in this previous calculations.
The Born effective charge on the $j$th atom is
$e{\bf Z}^*_{j\alpha\beta}=F_{j\beta}/E_{\alpha}$.
Traditionally, the Born effective charge is obtained via
$e{\bf Z}^*_{j\alpha\beta}=\Omega (\partial P_{\alpha}/\partial u_{j\beta})|_{{\bf E } =0}$
where $u_j$ is the displacement of the $j$th atom due
to the optical phonon mode, and is usually calculated by the finite difference approach.
Therefore, the finite field calculations have an advantage of being simple and 
less CPU-time consuming especially for complex systems like BN-NTs.

We consider a number of representative BN-NTs with a range of
diameters from all three types of BN-NTs, namely, the zigzag
[($n$,0), $n$ = 5, 6, 8, 12, 16], armchair [($n$,$n$), $n$ = 3, 4, 6, 8, 12], 
and chiral (4,2), (5,2) BN-NTs. 
For comparison, we also study the dielectric properties of bulk BN in 
the cubic, wurtzite and hexagonal structures (denoted as c-BN, w-BN and h-BN, respectively)
as well as a single honeycomb BN sheet.
Our {\it ab initio} calculations were performed using accurate
projector augmented-wave (PAW) 
method~\cite{blo94}, as implemented in the QMAS (Quantum MAterials Simulator)
package \cite{ish06}.
A supercell geometry was adopted so that the nanotubes are aligned in a
square array with the closest interwall distance between adjacent nanotubes being
9 \AA.  The isolated BN sheet is simulated
by a slab-supercell approach with an inter-sheet distance of 15 \AA. 
A plane-wave cut-off of 35 Ry was used throughout.
For BN-NTs, a $2\times2\times n$ mesh in the first Brillouin zone is used throughout. 
For the achiral nanotubes, $n = 8$ and for the chiral nanotubes, $n=6$. 
The ideal nanotubes were first constructed by rolling-up a honeycomb BN sheet. 
Their atomic positions and lattice constants were then fully relaxed 
%by a conjugate gradient technique. Theoretical equilibrium nanotube structures were obtained
until all the atomic forces and the stress were less 
than 4.0$\times10^{-5}$ Ry/Bohr
and 1.0$\times10^{-6}$ Ry/Bohr$^3$, respectively. 
The same force and stress convergence criteria were used in the
finite-field calculations. The small finite-field used is 
0.001 atomic unit \cite{uma02}.
The calculated field-induced polarization and Born effective
charge for the field perpendicular to the tube axis were checked with
the usual saw-tooth potential method which cannot be applied for the
field parallel to the tube axis, and the results from both approaches
agree with each other very well (up to the third digit).
Note that the unit cell
volume $\Omega$ is not well defined for nanotubes and the single BN sheet.
Therefore, like the previous calculations~\cite{guo04,guo05}, we used an
effective unit cell volume of the nanotubes and the BN sheet rather than 
the volume of the supercells
which is arbitrary. The effective unit cell volume for a nanotube and the
BN sheet is given, respectively, by
%$\Omega = \pi[(D/2+d/2)^2-(D/2-d/2)^2]T = \pi DdT$ and by
$\Omega = \pi DdT$ and by $\Omega = Ad$ where $d$ is the thickness of the
BN sheet which is set to the interlayer distance of $h$-BN.
%(3.28 \AA~\cite{guo05}). 
$D$ and $T$ are the diameter and length of 
translational vector of the nanotube~\cite{guo05}, respectively.  
$A$ is the area of the surface unit cell of the BN sheet.
Furthermore, in the supercell approach to a single wire
(an isolated slab), the periodic dipole images can give rise to
a substantial depolarization field which not only affect the transverse
component of the field-induced properties but also decrease rather
slowly as the supercell dimension increases\cite{koz06}. Therefore,
the transverse dielectric constant, electric polarizability and
Born effective charges presented below, have been corrected for 
this artificial depolarization field 
via one-dimensional Clausius-Mossotti  relation\cite{tob04} for the
single BN sheet and two-dimensional Clausius-Mossotti  relation\cite{wir05,koz06}
for the BN-NTs. We have checked numerically that the transverse component
of these physical properties are independent of the supercell dimension used
(within 1 \%) for the single BN sheet and also the (5,0) BN-NT. 

\section{Results and Discussion}

\begin{table}
\caption{Lattice constants ($a$,$c$), static dielectric constant ($\epsilon$), 
and Born effective charge ($Z^*$) of bulk BN in cubic, wurtzite and hexagonal structures
(denoted as c-BN, w-BN and h-BN, respectively) as well as the isolated BN sheet.
The available experiment values are listed in brackets.
}
\begin{ruledtabular}
                                                                                                
\begin{tabular}{ccccc}
     & c-BN & w-BN & h-BN & BN sheet \\ \hline
$a$  & 3.582 (3.615$^a$) & 2.524 (2.558$^a$) & 2.491 (2.50$^b$) & 2.489  \\
$c$  & -             & 4.177 (4.228$^a$) & 6.431 (6.65$^b$) & -      \\
$\epsilon_{\infty}^{aa,bb}$  & 4.47 (4.5$^c$)  & 4.37 (4.95$^d$)  & 4.69 & 4.70\\
$\epsilon_{static}^{aa,bb}$  & 6.79 (7.1$^c$)  & 6.25 (6.85$^d$)  & 5.50 & 7.20, 5.80 \\
$\epsilon_{\infty}^{cc}$      &             & 4.49 (4.10$^d$)     & 2.88 & 1.62      \\
$\epsilon_{static}^{cc}$      &             & 6.60 (5.06$^d$)     & 3.51 & 1.65       \\
$Z^*_{a(b)}$ & 1.88 (1.98$^e$)&  1.82  & 2.71  & 2.70     \\
$Z^*_c$     &            &  1.91  & 0.86 & 0.25
\end{tabular}

\end{ruledtabular}
\footnotetext[1]{Taken from Ref. \onlinecite{som74}.} 
\footnotetext[2]{Taken from Ref. \onlinecite{cap96}.} 
\footnotetext[3]{Taken from Ref. \onlinecite{gie67}.} 
\footnotetext[4]{Taken from Ref. \onlinecite{gei66}.} 
\footnotetext[5]{Taken from Ref. \onlinecite{san83}.}
\end{table}

\subsection{Bulk BN structures and single BN sheet}

The results for the bulk BN structures which are summarized in Table I.
%The theoretical lattice constants for h-BN ($a = 2.491$ \AA) and 
%for the BN sheet ($a = 2.489$ \AA) are nearly identical. 
All the theoretical lattice
constants of all bulk BN structures agree rather well (within 1.5 \%) with the 
experimental values (see Table I), except $c$ of h-BN which is 3.3 \% smaller
than the experimetal value.
Interestingly, the static dielectric constants and Born effective charges
of c-BN and w-BN are very similar, and the anisotropy for w-BN is rather
small (Table I). This is due to the fact that although the layer stacking
along the $c$ direction of the wurtzite structure and the one along [111] direction
of zincblende structure are different, the local tetrahedral atomic configuration
is common to both structures.
The calculated dielectric properties for c-BN and w-BN  
agree rather well (within 4.0 \%) with the
available experimental values (see Table I).
In contrast to w-BN, the anisotropy in the dielectric properties of h-BN is rather
pronounced. For example, the Born effective charge for the field along the
$c$-axis ($Z_c$) is only about 1/3 of that for the field perpendicular to
the $c$-axis ($Z_{a(b)}$) due to weak covalency along the $c$-axis. 
Here {\bf a} denotes one of the two in-plane primitive vectors (i.e., the vector
connecting the two nearest neighoring B(N) atoms) and {\bf b} is the vector that
is perpendicular to {\bf a}.

It is interesting to compare the last two columns (h-BN and BN sheet) in Table I.
As h-BN is a one-dimensional stacking of BN sheets with rather large separation
between neighboring sheets, it is natural to expect that both systems
may have similar in-plane properties, as is actually seen in Table I except
the static dielectric constant. While there is no in-plane anisotropy for the clamped
ion dielectric constant and Born effective charge for both h-BN and BN sheet, only
the static dielectric constant of BN-NT shows strong in-plane anisotropy. The reason
for this is that lattice strain is induced in a BN sheet under the in-plane 
electric field. The mode of strain is schematically shown in Fig. 1, where the
maginitude of strain is much exaggerated. The strain mode is clearly anisotropic
with respect to the direction of electric field and induces the
anisotropy in the in-plane static dielectric constant for BN sheet.
However, such a lattice strain cannot be induced in h-BN because of the alternating
array of B and N along the $c$-axis. The B and N arrangement of a given sheet
is reversed in the neighboring sheets so that the strain induced by electric field in the
neighboring BN sheets cancels out.

\begin{figure}
\includegraphics[width=8cm,trim= 0.cm 3.cm 2.cm 0.cm]{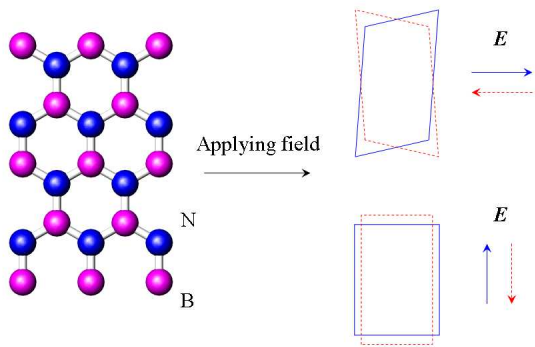}
\caption{\label{fig1} (Color online) (a) Atomic arrangement of a single BN sheet and
(b) Schematic illustlation of the mode of deformation of BN sheet under the in-plane
electric field.  Note that the magnitude of deformation is exaggerated.}
\end{figure}

\begin{figure}
\includegraphics[width=8cm]{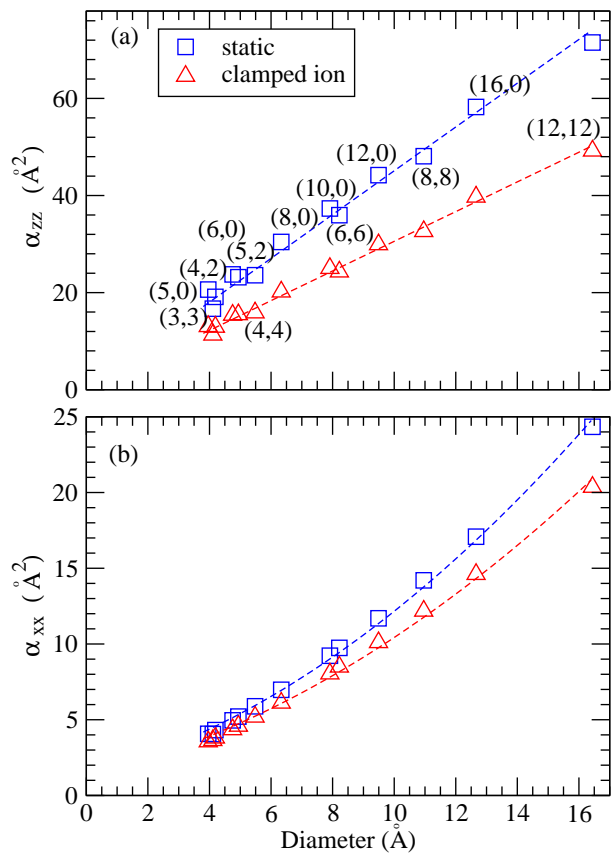}
\caption{\label{fig2} (Color online) (a) Longitudinal ($\alpha_{zz}$) and (b) transverse
($\alpha_{xx}$) polarizabilities of BN-NTs as a function of tube diameter ($D$). The dashed
lines in (a) are the straight lines fitted to the calculated polarizabilities (symbols), and
in (b) are the function of $b(D+\delta)^2$ fitted to the calculated values.}
\end{figure}

%\section{Dielectric constant and electric polarizability}

%\begin{table}
%\caption{Static dielectric constant ($\epsilon_{static}$), 
%electric polarizability ($\alpha$) and
%Born effective charge ($Z^*$) of the BN nanotubes.
%}
%\begin{ruledtabular}
%                                                                                                
%\begin{tabular}{c c c c}
% tube & $\epsilon^{xx}_{static}$,$\epsilon^{zz}_{static}$
%      & $\alpha_{xx}$, $\alpha_{zz}$
%      & $Z^*_{x}$, $Z^*_{y}$, $Z^*_{z}$ \\
%      &     &  (\AA$^2$) & (e) \\ \hline
% (5,0)  & 2.25, 7.36  & 4.07, 20.64 &  0.34, 0.93, 2.87 \\
% (6,0)  & 2.27, 7.10  & 4.96, 23.76 &  0.32, 0.99, 2.83 \\
% (8,0)  & 2.34, 6.86  & 6.98, 30.43 &  0.30, 1.10, 2.78 \\
% (12,0) & 2.50, 6.68  & 11.69, 44.20 &  0.28, 1.30, 2.74 \\
% (16,0) & 2.65, 6.61  & 17.08, 58.24 &  0.27, 1.47, 2.72\\
%%(20,0) &             &              &                  \\
% (3,3)  & 2.20, 5.95  &  4.05, 16.69 &  0.37,  0.95, 2.56 \\
% (4,4)  & 2.31, 6.25  &  5.89, 23.59 &  0.33,  1.06, 2.62 \\
% (6,6)  & 2.45, 6.33  &  9.75, 35.96 & 0.29, 1.25, 2.66 \\
% (8,8)  & 2.58, 6.35  &  14.20, 48.09 & 0.28, 1.40, 2.67  \\
% (12,12)& 2.81, 6.31  &  24.34, 71.52 & 0.27, 1.61, 2.69 \\
% (4,2)  & 2.26, 6.57  & 4.32, 19.14  & 0.36, 0.95, 2.65 \\
% (5,2)  & 2.29, 6.72  & 5.21, 23.17  & 0.33, 1.01, 2.70 \\
% (6,0) bundle & 3.79, 7.04 & 10.59, 23.60 & 0.54, 1.90, 2.86 
%%(12,0) bundle &            &              &                
%\end{tabular}
%\end{ruledtabular}
%\end{table}

\subsection{BN-NT}

\subsubsection{Longitudinal dielectric constant and electric polarizability}

The calculated static dielectric polarizability $\alpha_{zz}$ and dielectric
constant $\epsilon_{static}$ of the BN-NTs are plotted as a function of diameter
($D$) in Figs. 2 and 3, respectively. Fig. 2(a) shows that on average, 
both the full and clamped ion (electronic contribution only)
polarizabilities $\alpha_{zz}$ per unit length for the electric field parallel
to the tube axis ($E\parallel \hat{z}$) is roughly
proportional to the tube diameter, %[Fig. 1(a)], 
i.e., $\alpha = aD$,
resembling the behavior of a thin dielectric cylinderical shell.
%being independent of chirality and electronic structure.
This behavior is in strong contrast with the latest finding of $\alpha_{zz}\sim D^2+C$
for the CNTs from the density-functional perturbation theory
(DFPT) calculations~\cite{koz06} and also with the recent finding of
from the IPLRT %independent-particle linear response theoretical 
calculations\cite{guo04}.
Because the number of atoms on a BN-NT per unit length is
proportional to $D$, this result indicates that every atom
on the BN-NTs has nearly the same static polarizability, as would
be the case for insulators.
The slope of $\alpha_{zz}$ is 
$a_z$ = 4.57 \AA$ $ and $a_{z,0}$ = 3.08 \AA$ $ for the
full and clamped ion polarizabilities, respectively.
Note that the slope $a_{z,0}$ is nearly identical 
to the previous IPLRT calculations (3.04 \AA\cite{guo05}, 3.13\cite{wir05}) 
indicating that local field effects
for $E\parallel \hat{z}$ are indeed negligible.
However, when the ionic contribution is included, as in the
present calculations, the electric polarizability $\alpha_{zz}$ is 
significantly increased,
by $\sim$ 50 \%, demonstrating the importance of the ionic contribution.

There is clear chirality dependence in $\alpha_{zz}$ 
and $\epsilon_{zz}$, as shown in Fig. 2(a) and Fig. 3. 
We first discuss the clamped ion dielectric constants which are
denoted by inverse triangles in Fig. 3 to simplify the problem. For a given
diameter of BN-NT, zigzag BN-NTs have larger dielectric constant than armchair ones and
the chiral ones are intermediate. The chirality dependence is strong for smaller 
diameters and tends to vanish as the diameter increases. The longitudinal dielectric
constant of clamped ion case clearly converges to 4.7 which is the clamped ion
dielectric constant of a single BN sheet, which is expected to be the case. The
strong chirality dependence of clamped-ion longitudinal dielectric constant
comes from the chirality dependence of the band gap. It was reported in
Ref. \onlinecite{guo05} that the band gap of   
the zigzag BN-NTs would be reduced substantially as the tube diameter 
decreases, while that of the armchair BN-NTs depends 
on the diameter only slightly. It is known that the systems with a 
smaller band gap would have a larger dielectric constant. 
For static longitudinal dielectric constant, the present calculation takes
account of only the relaxation of atomic positions with a fixed unict cell.  Due to
the difficulty in practical calculations, the shear deformation of the unit cell
as observed in a single BN sheet (Fig. 1) is not taken into accout.  Therefore
the behavior of the static longitudinal dielectric constant shown by triangles in
Fig. 2 is basically the same as those of clamped ion case.  The large diameter limit
of static dielectric constant is about 6.4 which is close to the value for a single
BN sheet obtained within the same constraint of atomic relaxation.
 
\begin{figure}
\includegraphics[width=8cm]{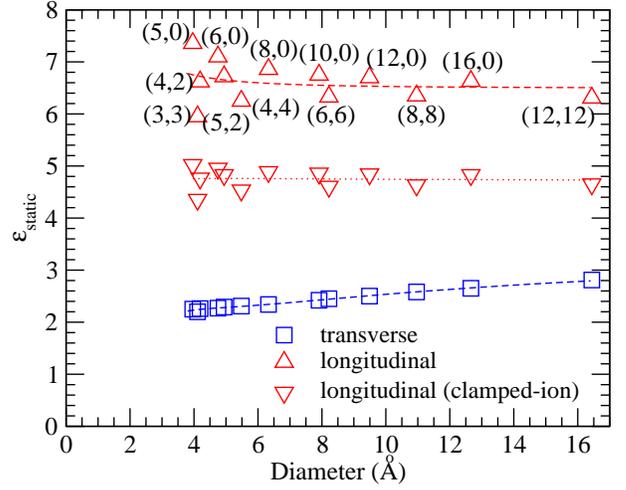}
\caption{\label{fig3} (Color online) Full static dielectric constant of BN-NTs as a function
of tube diameter. Each dashed line is the curve fitted to the corresponding calculated 
dataset. Longitudinal clamped-ion dielectric constants of BN-NTs are also plotted
for comparison.}
\end{figure}

\subsubsection{Transverse dielectric constant and electric polarizability}

We begin our discussion with rather simple features. First, the transverse
quantities (in Fig. 2(b) and Fig. 3) do not show appreciable chirality dependence
for the whole range of diameter.  Second, the ionic contribution is rather small, 
being only $\sim$ 15\% of the electronic contribution in the
polarizability [Fig. 2(b)]. Third, the large diameter limit of transverse
dielectric constant may be given by the circular average of the response of each part
of the BN-NT surface, which may be approximated by $0.5({\epsilon_{static}^{aa,bb}}
+ {\epsilon_{static}^{c}})$ with quantities for a single BN sheet.  As we do not
take account of the shear strain for BN-NT, $\epsilon_{static}^{aa,bb}$ is about
6.4 and the converged value of the squares (in Fig. 3) in the large diameter limit 
may be about 4.0.
The most surprising fact in the transverse quantities is that the transverse 
polarizability $\alpha_{xx}$ in Fig. 2(b)
exhibits a rather different behavior\cite{guo05,wir05}. In fact, $\alpha_{xx}(D)$ can be nicely
fitted to $\alpha_{xx} = b(D+\delta)^2$, as can be seen 
from Fig. 2(b). This indicates that as far as the transverse dielectric
properties are concerned, a BN-NT behaves 
like a thin (nonideal) conducting cylinderical shell\cite{krc03}
(or a system of electrons moving freely on a cylindrical shell\cite{ben95})
with an effective diameter of $D+\delta$.
The parameters $b$ and $\delta$ obtained by the fitting are, respectively,
0.0629 and 4.13 for the full polarizability, and 0.0523 and 4.40 for the
clamped ion polarizability. $\delta$ is slightly larger than the effective thickness
($d\sim 3.3$\AA ) of the BN sheet mentioned above. The screening factor of 
the inner electric field ($E_{in}$) by a BN-NT is 
$E_{out}/E_{in} = 1/(1-8b) \approx 2$ ($E_{out}$ is the
applied field outside the BN-NT) when the full
response of the BN-NT is considered. Note that for a long classical dielectric
cylindrical shell, the inner electric field {\bf E}$_{in}$ is also uniform and along
{\bf E}$_{out}$. However, 
$E_{in} = 4E_{out}\epsilon/[(\epsilon + 1)^2 - (\epsilon - 1)^2(D+d)^2/(D-d)^2]$
would be enhanced rather than screened. For example, for the (8,8) BN-NT, 
the screening factor is 0.4 if $\epsilon_{static}^{xx}$ (2.58) [for the (8,8) BN-NT] 
is used or 0.8 if $\epsilon_{static}^{cc}$ (1.65) [for the BN sheet] is used. 
To see whether this phenomenological interpretation in term of a thin conducting
cylindrical shell is an accidental coincidence or not, we plot the calculated
electric potential on a plane perpendicular to the tube axis of the (8,8) BN-NT
in Fig. 4(a) and also calculate the total electric field along the applied 
electric field line going through the center of the BN-NT [Fig. 4(b)]. Fig. 4 clearly
shows that the electric field inside the BN-NT is indeed uniform and considerably reduced.
The screening factor from Fig. 4(b) is 2.08 but not infinite as would be expected from
a perfect conducting cylindrical shell. We also plot the electric field for some other
BN-NTs and find the same situation. Note that both the metallic and semiconducting
CNTs exhibit the same behavior with a larger screening factor of 4.4.\cite{koz06} 
Therefore, the transverse polarizability from the present finite field calculations
reveals a qualitatively different behavior, 
when compared with previous IPLRT calculations,~\cite{guo05,wir05}
demonstrating the importance of taking the local field effects into account
in this case. 

\begin{figure}[tb]
\includegraphics[width=8cm]{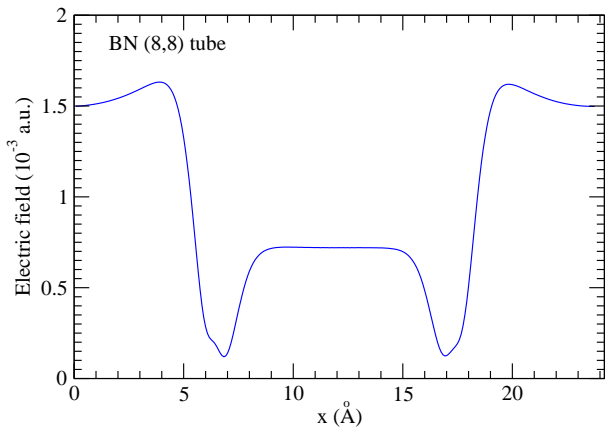}
\caption{\label{fig4} (Color online) Upper panel: Potential contour plot on a xy-plane perpendicular to the tube axis.
Lower panel: The electric field along the line parallel to the $x$-axis and going through
the center of the BN-NT in the xy-plane shown in the upper panel.
}
\end{figure}

\begin{figure}[tb]
\includegraphics[width=8cm]{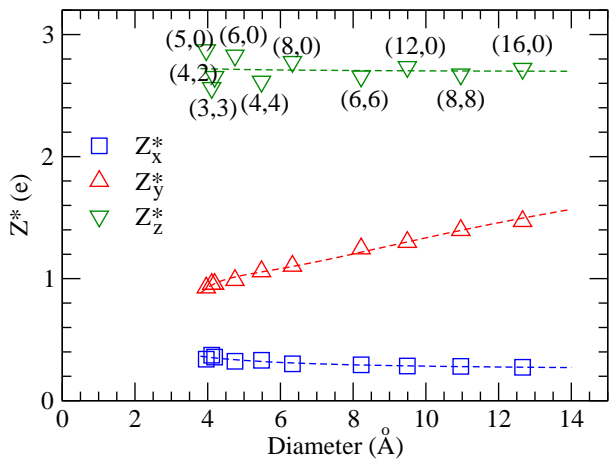}
\caption{\label{fig5} (Color online) Three principal Born effective charges
of BN-NTs as a function of tube diameter. Each dashed line is the 
curve fitted to the corresponding calculated dataset.}
\end{figure}

\subsubsection{Born effective charges}

To obtain the Born effective charges, 
we perform for each BN-NT
three self-consistent finite-field 
calculations for the field applied along three Cartesian
coordinate axes, respectively. A Born effective charge tensor for each atom is then
constructed from the field-induced atomic forces. 
These Born effective charge tensors are then diagonalized to give three principal
components ($Z^*_x,Z^*_y,Z^*_z$). Interestingly,
it turns out that $Z^*_x,Z^*_y$ are for the transverse fields perpendicular to and tangential
to the local tube surface of the ion concerned, respectively, while
$Z^*_z$ corresponds to the field along the tube axis.
The calculated principal Born effective charges on
all the BN-NTs studied here are displayed
as a function of the tube diameter in Fig. 5. 

Figure 5 shows that the medium component ($Z^*_y$) 
of the BN-NTs exhibits a pronounced monotonic dependence on the tube 
diameter (but independent of chirality) while the small component 
($Z^*_x$) is almost a constant. In contrast, the large component ($Z^*_z$) 
shows a clear chirality dependent oscillation superimposed on the mean value of
2.70 e (the in-plane components of the Born effective charge of the single
BN sheet) (see Fig. 5 and Table I).
These interesting diameter-dependence of $Z^*_y$ and
chirality-dependence of $Z^*_z$ suggest that one could
determine both the diameter and chirality of a BN-NT by measuring 
its Born effective charges
using, e.g., Raman and infrared spectroscopies.
Characterization of the chirality and diemeter of a BN-NT is still a great
challenge.

%The large Born effective charge $Z^*_z$ is also roughly a constant, 
%though it clearly oscillates around the mean value of $\sim$ 2.7 e in the
%region of small diamters (Fig. 2). 
Interestingly, as the tube diameter becomes large, $Z^*_x$ and $Z^*_y$ 
tend to approach to the $Z^*_c$ and $Z^*_{a(b)}$ components of the Born effective 
charge of the single BN sheet for the field 
perpendicular and parallel to the sheet plane (Table I), respectively.
This is perhaps not surprising because the observed monotonic diameter dependence
of the $Z^*_x$ and $Z^*_y$ merely reflect the degree of the local curvature of the BN-NTs
with a finite diameter.

\section{Concluding Remarks}

To summarize, we have carried out a systematic {\it ab initio} 
study of the static dielectric properties of the BN-NTs
within density functional theory with the LDA plus 
the latest finite-field electric-enthalpy theory.
Specifically, the full static dielectric constant, electric
polarizability and Born effective charges of a number of BN-NTs as 
well as the single graphitic
BN sheet and bulk BN structures have been calculated.
The calculated lattice constants, dielectric constants and Born effective charges
for the bulk BN structures are in good agreement with available experiments.
We find that the ionic contribution
to the static dielectric response of BN nanotubes (BN-NTs) is
substantial and that a clear chirality-dependent oscilation is imposed
on a linear relation between the longitudinal electric polarizability
and the tube diameter, as for a thin dielectric cylinderical shell. 
We also find that the calculated features of the transverse
dielectric response of the BN-NTs resemble that of a thin conducting cylindrical 
shell, with a screening factor of 2 for the inner electric field.
Our calculations also show that for each BN-NT, the medium principal component
of Born effective charge tensor
has a pronounced dependence on diameter (but independent
on chirality) and the large principal componet exhibits a chirality dependence
(but nearly independent on diameter), suggesting that 
a BN-NT could be characterized by measuring its
Born effective charges using, e.g., Raman and infrared spectroscopies.
%Interwall interaction in a BN-NT bundle is found to have almost no effect
%on the longitudinal dielectric response of BN-NT, though it enhances considerably
%the transverse dielectric constants, electric polarizabilities and Born effective
%charges.

%\section*{Acknowledgments}

\begin{acknowledgements}
We thank M. Kohyama and S. Tanaka for their helps 
in developing the QMAS code.
G.Y.G gratefully acknowledge a guest researcher-ship from the AIST of Japan.
This work is partly supported by the Next Generation 
Supercomputing Project, Nanoscience Program and also partly by 
Grant-in Aids for Scientific Research in Priority Area 
"Anomalous Quantum Materials", both from MEXT, Japan and
also by the National Science Council,
Ministry of Economic Affairs (93-EC-17-A-08-S1-0006) and
NCTS of Taiwan.
\end{acknowledgements}

\end{document}